\documentclass{article} % For LaTeX2e
\usepackage{iclr2021_conference,times}

\usepackage{amsmath,amsfonts,bm}

\def\eqref#1{equation~\ref{#1}}
\def\1{\bm{1}}

\DeclareMathAlphabet{\mathsfit}{\encodingdefault}{\sfdefault}{m}{sl}
\SetMathAlphabet{\mathsfit}{bold}{\encodingdefault}{\sfdefault}{bx}{n}

\newcommand{\R}{\mathbb{R}}

\usepackage[hidelinks,colorlinks,linkcolor={red!80!black},citecolor={blue!60!black}, urlcolor={blue!60!black}]{hyperref}
\usepackage{url}

\usepackage[mathletters]{ucs}
\usepackage[utf8x]{inputenc}
\usepackage{pdfcomment}
\newcommand{\refequ}[1]{Eq.~(\ref{equ:#1})}

\newcommand{\reffig}[1]{Figure~\ref{fig:#1}}

\providecommand{\D}{}

\providecommand{\R}{}
\providecommand{\S}{}

\providecommand{\V}{}

\providecommand{\p}{}

\providecommand{\x}{}

\providecommand{\z}{}
\usepackage{tabularx}
\usepackage{booktabs}
\usepackage{makecell}
\usepackage[table]{xcolor}
\definecolor{white}{rgb}{1,1,1}
\definecolor{lightbluishgrey}{rgb}{0.76471,0.84824,0.91647}

\usepackage{subcaption}
\usepackage{wrapfig}
\usepackage{graphicx}

\usepackage{listings}
\usepackage{layouts}
\newcommand{\alec}[1]{{\pdfcomment[color=blue]{Alec: #1}}}
\newcommand{\tom}[1]{{\pdfcomment[color=green]{Tom: #1}}}
\newcommand{\derek}[1]{{\pdfcomment[color=red]{Derek: #1}}}

\renewcommand{\alec}[1]{}
\renewcommand{\tom}[1]{}
\renewcommand{\derek}[1]{}

\newcommand{\SDF}{g}
\renewcommand{\D}{\mathcal{D}}
\renewcommand{\S}{\mathcal{S}}
\renewcommand{\V}{\mathcal{V}}
\renewcommand{\x}{\vec{x}}
\renewcommand{\p}{\vec{p}}
\renewcommand{\z}{\vec{z}}
\newcommand{\repName}{weight-encoded neural implicit}
\usepackage{sidecap}

\title{
  \centering
  On the Effectiveness of\\Weight-Encoded Neural Implicit 3D Shapes
}

\iclrfinalcopy

\author{Thomas Davies$^1$, Derek Nowrouzezahrai$^2$  \& Alec Jacobson$^1$  \\
$^1$Department of Computer Science, University of Toronto\\
$^2$Centre for Intelligent Machines, McGill University \\
}

\begin{document}

\maketitle

\begin{abstract}
  A neural implicit outputs a number indicating whether the given query point in
  space is inside, outside, or on a surface.
  Many prior works have focused on \emph{latent-encoded} neural implicits, where
  a latent vector encoding of a specific shape is also fed as input.
  While affording latent-space interpolation, this comes at the cost of
  reconstruction accuracy for any \emph{single} shape.
  Training a specific network for each 3D shape, a \emph{weight-encoded} neural
  implicit may forgo the latent vector and focus reconstruction accuracy on the
  details of a single shape.
  While previously considered as an intermediary representation for 3D scanning tasks
  or as a toy-problem leading up to latent-encoding tasks, weight-encoded neural
  implicits have not yet been taken seriously as a 3D shape representation.
  In this paper, we establish that weight-encoded neural implicits meet the
  criteria of a first-class 3D shape representation.
  We introduce a suite of technical contributions to improve reconstruction
  accuracy, convergence, and robustness when learning the signed distance field
  induced by a polygonal mesh --- the \emph{de facto} standard representation.
  Viewed as a lossy compression, our conversion outperforms standard techniques
  from geometry processing.
  Compared to previous latent- and weight-encoded neural implicits we
  demonstrate superior robustness, scalability, and performance.

\end{abstract}

\section{Introduction} \label{introduction}

% Narrow in on topic: 3D shape representations and Neural Networks → Neural
% implicits.
%
%Representing a solid shape efficiently particularly challenging because while
%queries depend solely on the shape's 2D surface the shape's 3D embedding allows
%it to exhibit geometric and topological complexities not well captured by planar
%representations (e.g., images).
%
While 3D surface representation has been a foundational topic of study in the
computer graphics community for over four decades, recent developments in
machine learning have highlighted the potential that neural networks can play as
effective parameterizations of solid shapes.

The success of neural approaches to shape representations has been evidenced
both through their ability of representing complex geometries as well as their
utility in end-to-end 3D shape learning, reconstruction, and understanding and
tasks. 
These approaches also make use of the growing availability of user generated 3D
content and high-fidelity 3D capture devices, e.g., point cloud scanners.

%
%\alec{need to mention these here?}
%Early uses \alec{trigger?} of neural networks for 3D solids built direct analogs of convnets for 3D images
%(e.g., voxel grids \cite{Maturana2015VoxNet}), fixed-size template meshes (e.g.,
%\cite{wang2018pixel2mesh}), or composites of image-based data (e.g.,
%\cite{groueix2018,williams2019}). 
%
For these 3D tasks, one powerful configuration is to represent a 3D surface $\S$ as
the set containing any point $\x∈ℝ³$ for which an implicit function (i.e., a
neural network) evaluates to zero:
\begin{equation}
  \label{equ:latent}
  \S := \big\{ \x ∈ ℝ³ | f_{θ}(\x;\z) = 0\big\},
\end{equation}
\begin{wrapfigure}[6]{r}{1.7in}
  \includegraphics[width=0.95\linewidth,trim=0mm 0mm 0mm 5.5mm]{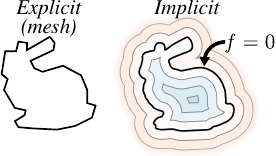}
  %\caption{}
\end{wrapfigure}%
where $θ∈ℝ^m$ are the network weights and $\z∈ℝ^k$ is an input latent vector
encoding a particular shape.
In contrast to the \emph{de facto} standard polygonal mesh representation which
\emph{explicitly} discretizes a surface's geometry, the function
$f$ \emph{implicitly} defines the shape $\S$ encoded in $\z$.
We refer to the representation in \refequ{latent} as a \emph{latent-encoded
neural implicit}.

\cite{Park_deepsdf} propose to optimize the weights
$θ$ so each shape $\S_i∈ \D$ in a dataset or shape distribution $\D$ is
encoded into a corresponding latent vector $\z_i$.
If successfully trained, the weights $θ$ of their \textsc{DeepSDF} implicit function $f_θ$ can be said to
generalize across the ``shape space'' of $\D$.
As always with supervision, reducing the training set from $\D$
will affect $f$'s ability to generalize and can lead to overfitting. 
Doing so may seem, at first, to be an ill-fated and uninteresting idea.

Our work considers an extreme case -- when the training set is reduced to a
single shape $\S_i$.
We can draw a simple but powerful conclusion: in this setting, one can
completely forgo the latent vector (i.e., $k=0$). 
From the perspective of learning the shape space of $\D$, we can
``purposefully overfit'' a network to a single shape $\S_i$:
\begin{equation}
 \S_i := \big\{x∈ℝ³ | f_{θ_i}(x) = 0\big\},
\end{equation}
where $θ_i$ now parameterizes a \emph{weight-encoded neural implicit} for the single shape $\S_i$.

In the pursuit of learning the ``space of shapes,'' representing a single shape
as a weight-encoded neural implicit has been discarded as a basic validation check
or stepping stone toward the ultimate goal of generalizing over many shapes
(see, e.g., \citep{chen2018implicit_decoder,Park_deepsdf,SAL,SALpp}). Weight-encoded neural
implicits, while not novel, \emph{\textbf{have been overlooked}} as a valuable shape
representation beyond learning and computer vision tasks.
For example, the original \textsc{DeepSDF} work briefly considered -- and nearly immediately discards -- the idea of independently encoding each shape of a large collection:
\begingroup
\addtolength\leftmargini{-0.1in}
\begin{quote}
``\textit{Training a specific neural network for each shape is
neither feasible nor very useful.}''\\\hspace*{\fill}-- \cite{Park_deepsdf}
\end{quote}
\endgroup

We propose training a specific neural network for each shape and
will show that this approach is both feasible and very useful.

\begin{wrapfigure}[]{r}{2.25in}%
\begingroup
\newcommand{\NO}{{\leavevmode\color[HTML]{e87c17}$×$}}
\newcommand{\OFTEN}{{\leavevmode\color[HTML]{e87c17}$×$}/$\bullet$}
\newcommand{\YES}{$\bullet$}
\centering
\renewcommand{\arraystretch}{1.2}
\setlength{\tabcolsep}{5.5pt}
\rowcolors{2}{white}{lightbluishgrey}
  \vspace*{-0.2in}
\begin{tabular}{l c c c c c c c c c c c c c c c c c c}
\rowcolor{white}
  %Representation    & \makecell{Continuous \\Surface Visualization}& \makecell{Trivial Spatial \\ Querying \& CSG} & Compact & \makecell{Consistent \\ Vectorization} \\
                    & I& II & III & IV \\
\midrule                                                                                                           
  Point cloud       & \NO                                          & \NO                                           & \YES    & \OFTEN                                 \\
  Mesh              & \YES                                         & \NO                                           & \YES    & \NO                                    \\
  Regular grid      & \YES                                         & \YES                                          & \NO     & \YES                                   \\
  Adaptive grid     & \YES                                         & \YES                                          & \YES    & \NO                                    \\
  Neural implicit   & \YES                                         & \YES                                          & \YES    & \YES                                   \\
\bottomrule
\end{tabular}
\endgroup
\end{wrapfigure}
We establish that a weight-encoded neural implicit meets the criteria of a
first-class representation for 3D shapes ready for direct use in graphics and
geometry processing pipelines (see inset table)
%
%Weight-encoded neural implicits are 
%suitable for direct application across the entire graphics and geometry processing
%pipelines: e.g., for rendering, distance querying, and constructive solid geometry
%(CSG) modeling.
%
While common solid shape representations have some important features and miss
others, neural implicits provide a new and rich constellation of features.
Unstructured point clouds are often raw output from 3D scanners, but do not
admit straightforward smooth surface visualization (I).
While meshes are the \emph{de facto} standard representation, conducting signed
distance queries and CSG operations remain non-trivial (II).
Signed distances or occupancies stored on a regular grid admit fast spatial
queries and are vectorizeable just like 2D images, but they wastefully sample
space uniformly rather than compactly adapt their storage budget to a particular
shape (III).
Adaptive or sparse grids are more economical, but, just as 
meshes will have a different number of vertices and faces, adaptive grids will
different storage profiles and access paths precluding consistent data
vectorization (IV).

While previous methods have explored weight-encoded neural implicits as an
\emph{intermediary} representation for scene reconstruction (e.g., \citep{NERF})
and noisy point-cloud surfacing tasks (e.g., \citep{SAL,SALpp}), we consider
neural implicits as the \emph{primary} geometric representation. Beyond this
observational contribution, our technical contributions include a proposed
architecture and training regime for converting the (current) most widely-adopted 3D
geometry format -- polygonal meshes -- into a weight-encoded neural implicit
representation. 

We report on experiments\footnote
{
  Source code, data, and demo at our (anonymized)
  repo:
\hyperref%
[github.com/u2ni/ICLR2021]
{https://github.com/u2ni/ICLR2021}
}
with different architectures, sampling techniques, and
activation functions -- including positional encoding \citep{NERF} and
sinusoidal activation approaches \citep{SIREN} that have proven powerful in the
context of neural implicits.
Compared to existing training regimes, we benefit from memory improvements
(directly impacting visualization performance), stability to perturbed input
data, and scalability to large datasets.

Weight-encoded neural implicits can be treated as an efficient, lossy
compression for 3D shapes.
Increasing the size of the network increases the 3D surface accuracy (see
\reffig{ablation}) and, compared to standard graphics solutions for
reducing complexity (mesh decimation and storing signed distances
on a regular grid), we achieve higher accuracy for the same memory
footprint as well as maintaining a SIMD representation: $n$ shapes can be
represented as $n$ weight-vectors for a fixed architecture.

The benefits of converting an existing mesh to a neural implicit extends beyond compression: 
in approximating the signed distance field (SDF) of the model, neural implicits
are both directly usable for many 
tasks in graphics and geometry processing, and preferable in many contexts compared to traditional representations.
Many downstream uses of 3D shapes already mandate the conversion of meshes to
less accurate grid-based SDFs, due to the ease and efficiency of computation for
SDFs: here, neural implicits serve as a drop-in replacement.

\begin{wrapfigure}[6]{r}{2.25in}%
  \vspace*{-0.05in}
%\begin{table}
\begingroup%
\newcommand{\NO}{{\leavevmode\color[HTML]{e87c17}$×$}}%
\newcommand{\OFTEN}{{\leavevmode\color[HTML]{e87c17}$×$}/$\bullet$}%
\newcommand{\YES}{$\bullet$}%
\centering%
\renewcommand{\arraystretch}{1.2}%
\setlength{\tabcolsep}{5.5pt}%
\rowcolors{2}{white}{lightbluishgrey}%
\begin{tabular}{l c c c c c c c c c c c c c c c c c c}
\rowcolor{white}
  Encoding:      & Latent & Weight \\
\midrule                                                                                                           
  Interpolation: & trivial & non-trivial \\
  Scalability:   &    poor &   excellent \\
  Stability:     &    poor &   excellent \\
\bottomrule
\end{tabular}
\endgroup
%\end{table}
\end{wrapfigure}

Many works explore latent-encoding methods (e.g.,
\citep{Park_deepsdf,SAL,SALpp}), taking advantage of interpolation in latent
space as a (learned) proxy for exploration in the ``space of shapes''.
We show that this flexibility comes at a direct cost of other desirable
proprieties.
In particular, we show that latent-encoded neural implicits scale poorly as a
representation for individual shapes both at training and inference time.
Existing latent-encoded neural implicits are sensitive to the distribution of
training data: while they may perform well for large datasets of a limited
subclass of shapes (e.g., ``jet airplanes''), we show that training fails with
more general 3D shape datasets. Even within a class, existing methods rely on
canonical orientation alignment (see Figure \ref{fig:deepSDFRotation}) in order to alleviate some of this difficulty --
such orientation are notably (and notoriously) not present in 3D shapes captured
or authored \emph{in the wild} and, as a result, latent-encoded neural implicits
will fail to provide meaningful results for many real-world and practical shape
datasets.
Fitting latent-encoded neural implicits to each shape independently complicates
shape space interpolation, rendering it difficult though not impossible
\citep{metasdf}.
In contrast, weight-encoded neural implicits leverage the power of the neural
network function space without the constraints imposed by the requirement of
generalizing across shapes through latent sampling.

%\alec{already summarized covered above?}
%We experiment with a variety of architectures and activation functions. We
%propose unifying the treatment of sampling training queries in ℝ³ with the
%definition of the loss function, revealing how importance sampling can be
%leveraged. Rather than rely on visibility as a cue for occupancy, we introduce
%generalized winding numbers during sampling making our method robust to complex
%or malformed training data.
%
%As a demonstration of the effectiveness of this representation we reduce the
%38.85GB Thingi10k dataset \citep{zhou2016thingi10k} composed of a zoo of mesh file formats to a fully
%vectorized 10,000-long list of 64KB weight vectors. The homogeneity of this data
%facilitates large-scale processing of shapes and 3D ML, where the shape
%representation is itself a weight-encoding, harmonizing with concurrent
%meta-learning endeavors \cite{metasdf}.

\section{Method}
\label{sec:method}
Neural implicits soared in popularity over the last year.
While significant attention has been given to perfecting network architectures
and loss functions in the context of latent-encoding and point-cloud
reconstruction, there is relatively little consideration of the conversion
process from 3D surface meshes to weight-encoded neural implicits (e.g., both
\cite{Park_deepsdf} and \cite{SIREN} consider this task briefly).
We focus on identifying a setup to optimize weight-encoded neural implicits for
arbitrary shapes robustly with a small number of parameters while achieving a
high surface accuracy.
Once successfully converted, we consider how the weight-encoded neural implicit
representation compares to standard 3D model reduction techniques and how
choosing this representation impacts downstream graphics and geometric modeling
operations.

\subsection{Signed Distance Field Regression}
In general, the value of an implicit function $f$ away from its zero-isosurface
can be arbitrary. 
In shape learning, many previous methods have considered occupancy where $f(\x)$
outputs the likelihood of $\x$ being inside of a solid shape (and extract the
surface as the 50\%-isosurface)
\cite{NERF,OccupancyNet, Littwin_2019, chen2018implicit_decoder, Maturana2015VoxNet, Wang-2018-AOCNN}.
We instead advocate that $f$ should approximate the \emph{signed distance field}
(SDF) induced by a given solid shape. Learning properties aside (see, e.g.,
\citep{Park_deepsdf}), SDFs are more immediately useful in graphics and geometry
processing applications. 

Given a surface $\S = ∂\V$ of a volumetric solid $\V ⊂ ℝ³$, the signed distance field
$\SDF_\S : ℝ³ → ℝ$ induced by $\S$ is a continuous function of space that outputs the distance of a
query point $\x∈ℝ³$ modulated by $±1$ depending on whether $\x$ is inside or
outside of the solid:
\begin{align}
  \SDF_\S(\x) = \text{sign}_\S(\x) \; \min_{\p∈\S} \|\x-\p\|,
  \quad \text{where} \quad
  \text{sign}_\S(\x) = \begin{cases}
    -1 & \text{if $\x∈\V$,} \\
    \hphantom{-} 1 & \text{otherwise.}
  \end{cases}
  \label{equ:sdf}
\end{align}

Our goal is to regress a feed-forward network $f_θ$ to approximate the SDF
of a given surface $\S$:
\begin{equation}
\label{eqn:lSDF}
    f_\theta(\x) \approx \SDF_\S(\x).
\end{equation}
If successfully trained, the weights $θ∈ℝ^m$ encode a neural implicit
representing $\S$.

\subsubsection{Architecture}
Our proposed architecture is a feed-forward fully connected network with $N$
layers, of hidden size $H$. Each hidden layer has ReLU non-linearities, while
the output layer is activated by $\tanh$.

\begin{figure}[t!]
	\centering
	\includegraphics[width=\linewidth]{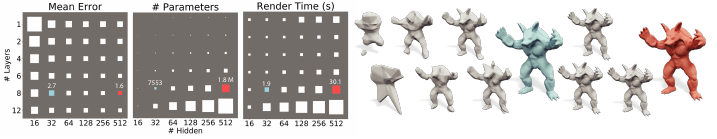}
	\vspace{-20pt}
	\caption{\label{fig:ablation}We visualize the role that varying the number of
		network layers and hidden layer sizes: (left to right) average
		reconstruction error, memory footprint and first-frame render
                time (DeepSDF, other setups, and our defaults in red, gray, and
                blue, respectively).
                }
	\vspace{-6 pt}
\end{figure}

Increasing the depth and width of this network will generally improve accuracy
but at the cost of increasing the memory footprint and, for example, the time
required to render the surface.
The \repName's rendered in Figures
\ref{fig:deepSDFRotation}, \ref{fig:thingi10k}, and \ref{fig:decimated} all
share a common architecture of just 8 fully connected layers with a hidden size
of 32 (resulting in just 7553 weights, or 59 kB in memory).
Through experimentation on a subset of 1000 meshes from Thingi10k
\citep{zhou2016thingi10k}, we find that this configuration yields a good balance
between reconstruction accuracy, rendering speed, and memory impact (Figure
\ref{fig:ablation}).
While maintaining acceptable surface quality, our default architecture has a
99\% reduction in number of parameters and 93\% speed up in ``time to render
first frame'' compared to the default weight-encoding architecture of
\citep{Park_deepsdf}.

Excited by the recent work exploring methods to overcome an MLP's bias to learn
low frequency signals faster, we performed experiments using both positional
encodings \citep{tancik2020fourier} and SIREN activations \citep{SIREN}.
Both perform well when the network architecture is sufficiently wide (e.g.,
$H>64$), but introduce surface noise with our more compact architecture.
See Appendix \ref{highFreq} for detailed experimental setup
and findings.

By increasing $N$ and $H$, our network could \emph{in theory}
\citep{hornik1989mlpuniversl} learn to emulate any arbitrary topology shape with
infinite precision.
In reality, like any representation, there are trade-offs.  The network
complexity can be increased over our base configuration for smaller surface
reconstruction error, or decreased for faster rendering speeds depending on the
application. A sample of geometries produced at a number of configurations can
be seen in Figure \ref{fig:ablation}.

\subsubsection{Integrated Loss $→$ Importance Sampling} \label{sampling}
Particularly choices of pointwise loss functions have been well explored by
previous papers \citep{Park_deepsdf,SAL,SALpp,gropp2020implicit,SIREN}, in our
experiments we find that a simple absolute difference $|f_θ(\x) - \SDF_\S(\x)|$
works well. 
Defining the total loss after the fact via \emph{ad hoc}
sampling (near-)surface sample process \citep{Park_deepsdf,SAL,SALpp} leaves an
unclear notion whether the total loss can be expressed as an integral and hides
possibly unwanted bias.
We focus instead on how to integrate this pointwise loss over space.

Sampling based on mesh vertices \citep{Littwin_2019,SIREN} reduces
accuracy in the middle of triangle edges and faces and introduces bias near regions
of the mesh (inset: Vertex) with denser vertex distributions regardless of the geometric
complexity or saliency of the region.
\begin{wrapfigure}[]{r}{0.5\linewidth}
	\centering
	%\vspace{-20pt}
	\includegraphics[width=\linewidth]{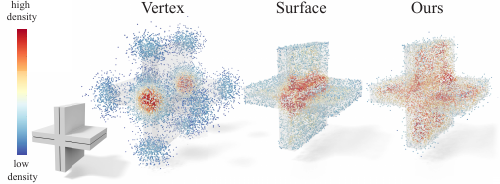}
	\vspace{-20pt}
	%\caption{\label{fig:importanceCompare}Visualizing sample density of samples drawn using:  svertex sampling with
	%	$\mathcal{N}(p,0.1)$ offsets~\citep{chen2018implicit_decoder}, surface sampling
	%	with $\mathcal{N}(p,0.01)$~\citep{Park_deepsdf,SIREN}, and our importance sampling
	%	approach with $w=e^{-30|SDF(p)|}$.  We color samples according to their
	%	density estimated with Gaussian kernel density, normalized by the most dense
	%	region from vertex sampling.}
\end{wrapfigure}

Similarly, sampling from Gaussians centered on the surface
\cite{Park_deepsdf,chen2018implicit_decoder,SAL,SALpp} will place over emphasis in regions of high
curvature, in thin solid/void regions (inset: Surface).

In contrast to \emph{ad hoc} samplings, we define the total loss
directly as an integral over space,
\begin{equation}
  \label{equ:total}
  L(θ) = ∫_{ℝ³} w(\x) \; |f_θ(\x) - \SDF_\S(\x)| \; d\x,
\end{equation}
where $w:ℝ³→ℝ_{≥0}$ is a non-negative weighting function with finite integral
over $ℝ³$.

Methods which randomly sample within a bounding box around a given shape
\citep{OccupancyNet,tancik2020fourier} can be understood as choosing $w$ to be
the characteristic function of the box. As \cite{Park_deepsdf} already observe,
this is wasteful if we care most that $f$ is accurate near the shape's surface
(i.e., where $\SDF_\S=0$).

We achieve this \emph{directly} --- without yet invoking sampling --- by
choosing $w$ exponentially as distance to $\S$ grows, specifically:
\begin{equation}
  w(\x) = e^{-β |g_\S(\x)|},
\end{equation}
where $β≥0$ can be adjusted from uniform sampling ($β=0$) to $β→\infty$ for
surface-only sampling. \alec{A reasonably balanced choice is $β=30$ for shapes
normalized to fit in the unit sphere.}

Attempting to sample space and measure the integrand of \refequ{total} directly
leads to many samples having little to no numerical effect during training. For
example, if $β=30$ and we consider a point unit distance away from the surface,
the weighting term itself closes in on machine double precision $w≈9e-14$.
By resisting the urge to prematurely sample until after we have written our
total loss function as an integral, we can instead apply \emph{importance
sampling} \citep{kahn1951} to construct a proportional approximation:
\begin{equation}
  L(θ) ≈ \sum_{\x ∈ \D_w} |f_θ(\x) - \SDF_\S(\x)|,
\end{equation}
where $\D_w$ is a distribution over $ℝ³$ with probabilities proportional to $w$.

We sample from $\D_w$ in practice via a simple subset rejection strategy.
Starting with a large (e.g., 10M) pool of uniform samples within a loose
bounding sphere around the shape, we re-sample (with replacement) a smaller
(e.g., 1M) subset with probability according to $w$.
Further improvements may be possible by incorporating advanced sampling patterns
\emph{à la} \cite{Xu2020}.

Compared to uniform sampling, weighting by our choice of $w$ leads to faster
convergence and reduced surface reconstruction when validating against a subset
of 1000 geometries from Thingi10k (96 epochs with surface error of 0.00231).
Compared to the sampling of \cite{Park_deepsdf}, we match convergence speed (86
epochs each) and demonstrate a $\approx 5\%$ improvement in surface error.

\begin{wrapfigure}[]{r}{0.5\linewidth}
	\centering
	\vspace{-5 pt}
	\includegraphics[width=\linewidth]{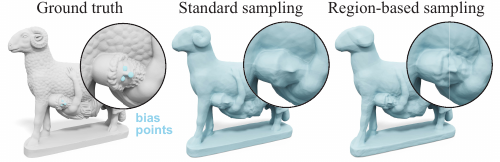}
	\vspace{-20 pt}
	%\caption{Our importance metric can be additionally weighted by distance
	%	from user specified regions. This weighting allows users to specify regions
	%	of interest (center; shown in red) yielding improved reconstruction accuracy
	%	(right) where desired.  \label{fig:pointBias}} 
\end{wrapfigure}
%Although the $\approx 5\%$ improvement in surface quality may be beneficial in
%some applications, 
Perhaps the most valuable property of our importance sampling
scheme to be its flexibility.

Our method has effectively removed all unintended
bias present in previous approaches, and enables complete user control on
\textbf{intended} bias to the sampling process.
The importance metric, $w(\x)$,
can be modified to explicitly bias importance toward regions of high curvature, minimum
feature size (emulating the hidden bias of \cite{Park_deepsdf}), or near user
annotations (see inset where $w(\x)$ is additionally scaled according to user selection). 
This flexibility allows for greater use of the network's capacity on areas
important to the user, without increasing overall network complexity or
radically changing the sampling protocol.

\subsection{Robust loss function for meshes in the wild}
The input $\S$ should be the boundary of a solid region $\V⊂ℝ³$; that is, a closed,
consistently oriented, non-self-intersecting surface.
%
%Strictly applying this pre-condition to common polygonal meshes found in use
%would severely limit the applicability of neural implicits.
%
Ignoring ``two-sided'' meshes that are not intended to represent the boundary of
a solid shape (e.g., clothing), many if not most meshes found online \emph{which
intend} to represent a solid shape would not qualify these strict
pre-conditions.
\cite{zhou2016thingi10k} 
\begin{wrapfigure}[15]{r}{0.5\linewidth}
	\vspace{-9pt}
	\centering
	\includegraphics[width=\linewidth]{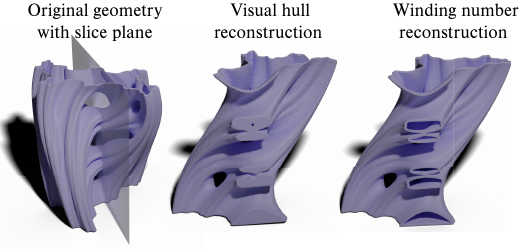}
	\includegraphics[width=\linewidth]{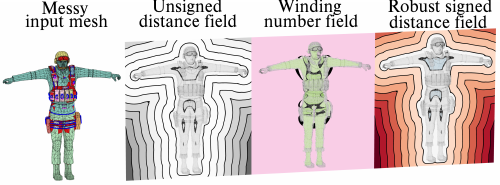}
	\vspace{-15pt}
	%\caption{Our approach to signing allows us to support converting non-manifold mesh, without sacrificing the true topology of the mesh. Unlike \cite{Park_deepsdf} visual hull method (middle), our method(right) maintains complex internal structures. Virtox (left) under CC BY.}
	\label{fig:visualHullFailure}
\end{wrapfigure}
observe that nearly 50\% of Thingi10k's solid 
models for 3D printing fail one criteria or another.
The failure point in terms of our equations so far is the definition of the
signing function $\text{sign}(\x)$  in \refequ{sdf} which relies on determining
whether a point $\x$ lies inside $\V$.

To determine insideness, previous approaches either require watertight inputs
\alec{citation needed}, use error-prone voxel flood-filling \citep{OccupancyNet}
or use inaccurate visual hulls as a proxy \citep{Park_deepsdf} (see
inset where visual hill signing can be shown to "close off" internal structure. Virtox (left) under CC BY. ).
Alternatively, \cite{SAL,SALpp} advocate for a loss function based on \emph{unsigned}
distances.
This introduces unnecessary initialization and convergence issues, that
can be avoided if we assume that the input mesh intentionally oriented to
enclose a solid region (as is the case for nearly all of Thingi10k), but may
suffer from open boundaries, self-intersections, non-manifold elements, etc.
Under these assumptions, the generalized winding number
\citep{jacobson2013robust} computes correct insideness for solid meshes and
gracefully degrades to a fractional value for messy input shapes
(see inset).
Using the tree-based fast winding numbers of \cite{Barill:FW:2018} and
a bounding volume hierarchy for (unsigned) distances, we can construct
our 1M-point sample set efficiently and optimize weights $θ$ for even the most
problematic meshes (see inset) in an average of 90 seconds per shape.

%
%Once trained, our neural implicit format can be treated as any other first class
%implicit representation.

\subsection{Efficient Visualization}

\alec{cite \cite{hart1986} and \cite{liu2019dist} here (previous use of sphere tracing for
latent-encoded neural implicit neural implicit)}

Our \repName\  representation can be treated as its classical counterpart (SDF)
and rendered efficiently using sphere-tracing \citep{hart1996sphere}.
Sphere tracing 
is a common technique for rendering implicit fields where rays are initialized
in the image plane and iteratively ``marched'' along by a step size equal
to the signed distance function value at the current location.
The ray is declared to have hit the surface when sufficiently close 
($<\epsilon$).
For more details, see Morgan McGuire's comprehensive notes at
\hyperref[casual-effects.com]{casual-effects.com}.

We trivially adapt traditional sphere-tracing by initializing the starting
position of each ray to be its first (if any) intersection with the similarity
transformed unit sphere, since all \repName s are normalized to lay within.
As rays of the image will converge different times, we employ a
dynamic batching method that composes batches of points for inference based on a
mask buffer which tracks rays that have converged to the surface or reached the
maximum number of steps. 
Local shading requires the surface normal at the hit point.
For SDFs, the unit normal vector is immediately revealed as the spatial gradient
(i.e., $∂f_{θ}/∂\x$).
This can be computed by finite differences or back propagation through
the network.

\section{Implementation and Results}

We implement {\repName} networks in Tensorflow (\cite{tensorflow2015-whitepaper}) with point sampling and mesh processing implemented in libigl (\cite{jacobson2016libigl}). We train our model for up to $10^2$ epochs and allow early stopping for quickly converging geometries. We use the ADAM optimizer (\cite{kingma2014adam}) with a fixed learning rate of $10^{-4}$. These settings generalized well across a wide range of geometries (see Figures \ref{fig:thingi10k} and \ref{fig:reconstructionHistograms}). 

\subsection{Surface Visualization and CSG} \label{sec:rendering}

We implement sphere-marching visualization and shading kernels in CUDA, using CUTLASS (\cite{cutlass}) linear algebra libraries for efficient matrix multiplication at inference-time. 

We achieve an average display frame rate of \textit{34 Hz} -- for the large subset of the Thingi10k dataset we visualize -- when rendering a single neural implicit at $512 \times 512$ resolution on an Nvidia P100 GPU. This a significant performance improvement over previous learnt implicit inference and display pipelines, attributed in large part to our compact representation. \cite{liu2019dist} present a specialized renderer capable of a 1 Hz display rate, however at the price of many conservative optimizations: these include overstepping along all rays by a factor of 50\%, increasing the convergence criteria (early stopping), and implementing a coarse-to-fine display strategy. While these additional optimizations could further improve our rendering speed (at the cost of reduced visual quality), we opt to rely on a simpler (and very efficient) standard sphere-marching SDF renderer.

Indeed, as our representation is a learnt representation of the SDF, we also inherit other important
\begin{wrapfigure}[]{r}{0.5\linewidth}
  \centering
  \vspace{-12.5pt}
  \includegraphics[width=\linewidth,trim=0 0 0 4mm, clip]{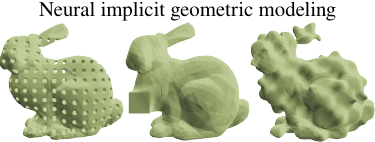}
%  \caption{\repName s easily allow for common interactive manipulations -- the model's predicted distances can be modified using boolean operations similar to any implicit field. See accompanying video for animation.}
    \vspace{-22.5pt}
  \label{fig:sdfops}
\end{wrapfigure}
benefits of traditional implicit function representations. Weight-encoded neural implicits admit robust shape manipulation and modification using constructive solid geometry operations (CSG) -- by directly modifying the inferred distance values (see inset and accompanying video). %We can smoothly interpolate between shapes (i.e., different weight vectors but the same architecture) by interpolating their weights. 
Weight-encoded neural implicits admit SIMD evaluation and, given their compactness, many neural implicits can be rendered in parallel at interactive rates on modern GPUs.% This opens up future avenues exploring the feasibility for animated neural implicit representation and the representation and rendering of \textit{entire scenes} using only neural implicits.

\subsection{Stability and Scale}

Training deep neural networks on large geometric datasets can be cumbersome and time consuming. For our \repName\  representation to be effective, we must be able to convert any 3D shape into its weight-encoded form in a reasonable amount of time. Due to our relatively simple
\begin{wrapfigure}[]{r}{0.5\linewidth}
	\vspace{-10pt}
	\includegraphics[width=\linewidth]{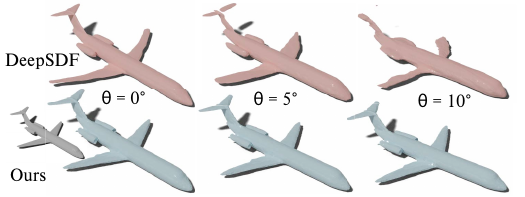}
	\vspace{-15pt}
	\caption{Unlike our representation, DeepSDF reconstruction quality degrades quickly for geometries not aligned to default, per-class orientations. See accompanying video for animation.}
	\vspace{-15pt}
	\label{fig:deepSDFRotation}
\end{wrapfigure}
base network architecture (8 layers of 32 neurons each) we find that we can overfit our model to any 3D shape \textbf{in 90 seconds}, on average. As this requires only 59 kB of memory, we can train many models/shapes concurrently on modern GPUs without approaching any practical memory limitations -- this ease of training is uncommon to other learning-based shape representations. Converting the entirety of the 10,000 models in the Thingi10k dataset \cite{zhou2016thingi10k} on an Nvidia Titan RTX only took 16 hours on a single GPU, or four hours on four Nvidia Titan RTX cards.% (see Figure \ref{fig:thingi10k} for results). 

\begin{wrapfigure}[]{r}{0.33\linewidth}
	\centering
	\vspace{-15pt}
	\includegraphics[width=\linewidth]{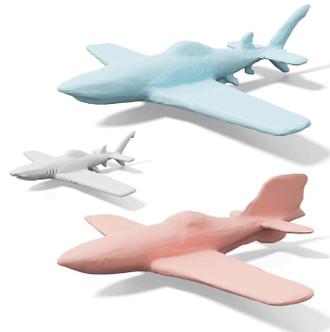}
	\vspace{-20pt}
	\caption{\label{fig:deepSDFFail} Latent-encoded SDFs (red) struggle to reconstruct ``unique" features (grey, plane's tail) despite training on a single class of objects (planes). Our representation (blue) does not. gpvillamil under CC BY.}
	\vspace{-15pt}
\end{wrapfigure}
Converting the Thingi10k dataset from mesh format to \repName\ format
reduces the overall storage from 38.85 GB to 590 MB -- a 1:66 compression rate. While a DeepSDF network \cite{Park_deepsdf} trained on the same dataset could compress this dataset to an impressive 7 MB footprint, the latent-decoded geometries it produces are of comparably lower quality. This comparison is representative, as Thingi10k is a real-world mesh dataset of objects obtained ``in the wild''. The dataset neither contains geometries aligned to a common frame of reference nor comprises objects nearing no semblance of inter-class categorization. These two properties make it difficult for any latent-encoded neural implicit network to converge to a reasonable result during training. 

We further support these claims using two experiments. First, we attempt to train DeepSDF on the Thingi10k dataset, and second we experiment with DeepSDF's ability to reconstruct shapes with slight perturbations from the shapenet \citep{chang2015shapenet} common shape orientation. Here, DeepSDF does not converge on the 10,000 model Thingi10k dataset, producing incoherent reconstructions when exploring the latent space of shapes it has learned. Moreover, if we further limit DeepSDF to training with a single class of objects, it is not able to reconstruct features on the tails of the inter-class distribution (inset, right). Secondly, we evaluate DeepSDF's ability to reconstruct geometries not aligned to the common orientation. Here, we retain single-class DeepSDF training and reconstruct the same input shape at orientations differing from the default (Figure \ref{fig:deepSDFRotation}). This test validates latent-encoding's reliance on having consistently aligned datasets, immediately precluding their use with large, real-world datasets.%is designed solely with the purpose of further validating our claims surrounding the reliance on consistently aligned datasets in DeepSDF -- such a reliance immediately discounts the application of DeepSDF to large, real-world datasets.

%\input{fig/sharpEdge} 
%As many of the geometries in Thingi10k dataset
%are organic "smooth" shapes, we also verify that our method is capable of maintaining sharp edges in reconstructions. We find that our network is able to recreate sharp edges (Fig. \ref{fig:sharpEdge} \& \ref{fig:failure}) with a high level of accuracy, despite not being specifically biased to do so. 

\subsection{Representation Compactness}

\begin{wrapfigure}[13]{r}{0.5\linewidth}
	\centering
	\includegraphics[width=\linewidth]{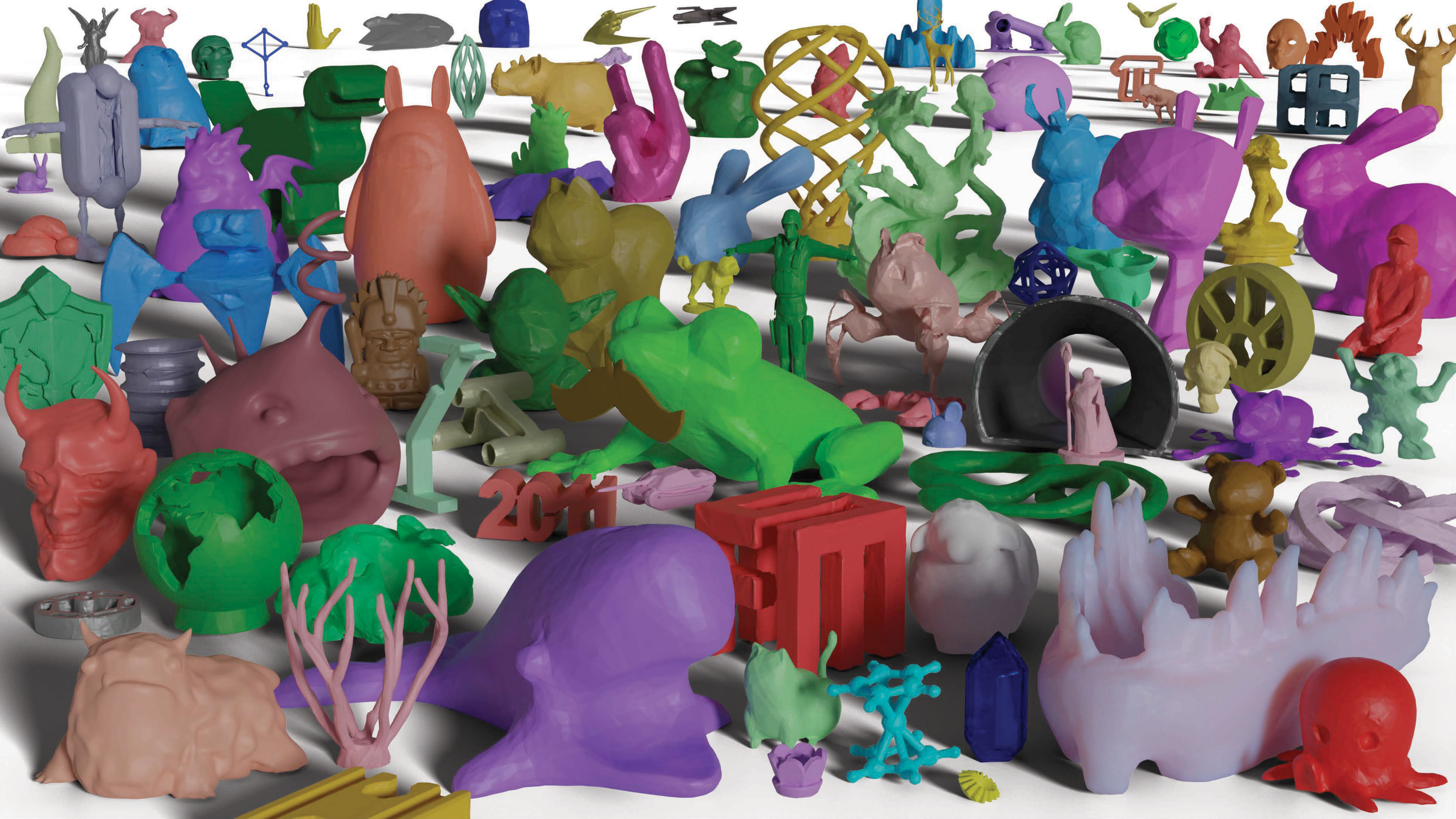}
	\vspace{-10pt}
	\caption{ Thingi10k models compressed to \textbf{59kB}, reducing the dataset from 38.85 GB to 590 MB. }
	\label{fig:thingi10k}
	\vspace{-10pt}
\end{wrapfigure}

All of the shapes in Figure \ref{fig:thingi10k} were rendered with \repName s generated using our base network architecture, resulting in a total of 7553 weights for each shape's implicit function. At just 59 kB of memory we find that our lightweight representation can capture complex geometric topologies at high resolution compared to uniform signed distance grids or adaptively decimated meshes with similar memory footprints.

The comparisons in Figure \ref{fig:decimation} use geometry converted to a
\repName\ in our base configuration, visualized next to the rendered result of a
uniformly sampled SDF grid with $20^3$ samples as well as with the original mesh
adaptively decimated \citep{garland1997surface} down to 7600 floats (i.e.,
vertex and face data). Compared to decimated meshes (our baseline non-uniform
format), we 
observe that weight-encoded neural 
implicits have similar surface
quality but with smoother reconstructions due to the continuous (versus
piecewise linear) nature of the implicit. Compared to SDFs stored on a grid (our
baseline uniform format) we observe far better quality at equal memory.
Furthermore, we notice that our approach better captures high frequency surface
detail compared to both these representations, often producing results that more
closely match the curvature of the original shape. 

\begin{wrapfigure}[]{r}{0.6666\linewidth}
	\vspace{-15pt}
	\includegraphics[width=\linewidth]{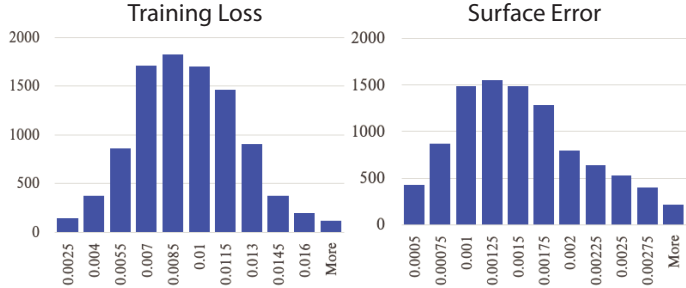}
	\vspace{-15pt}
        \caption{Loss and surface error distributions over the entirety of the
        Thingi10k dataset.}
	\vspace{-12pt}
	\label{fig:reconstructionHistograms}
\end{wrapfigure}

We measure our method's robustness by converting the Thingi10k (\citep{zhou2016thingi10k}) dataset and measuring the average surface error ($\textstyle {1}\big/{N}\sum_{i=1}^N |f_{\theta}(p_i)|$) and training loss. We report mean training loss for errors between the true and predicted SDF values at points sampled using our importance metric (Section \ref{sampling}). This surface error is the 
sum of errors at points along the shape's 0-isocontour. 
%
%\begin{equation} \label{eqn:surfaceError}
%    \text{Surface Error} = 
%\end{equation}
These metrics measure both the error at the surface and within the shape's bounding volume. Errors within the bounding volume decrease rendering performance and/or lead to hole artifacts in the shape during visualization. Surface errors are more evident after meshing the implicit SDF using, i.e., marching cubes. We sample $10^5$ surface points when measuring surface error, and compute loss against a training set of $1M$ points. We visualize results on the entire Thingi10k dataset in Figure \ref{fig:reconstructionHistograms}. We find that, at our base configuration, 93\% of the $10^5$ Thingi10k shapes reach a surface error below 0.003, and no model exceeds 0.01 (worst case of 0.0097; see Fig. \ref{fig:failure}).
 
\section{Limitations and Future Work}
Our default architecture will fail to satisfyingly approximate 
very topologically or geometrically complex shapes.
While increasing the size of the network will generally alleviate this (see
Figure \ref{fig:failure}), it would be interesting to consider cascading or adaptively
sized networks.
Our $L¹$ loss function encourages the network to match the \emph{values} of a
shape's signed distance field, but not necessarily its derivatives (cf.
\cite{gropp2020implicit,SIREN}). True SDFs satisfy an Eikonal equation
($|∂g/d\x|=1$) and this property is sometimes important for downstream tasks.
For future work, we would like to investigate whether Eikonal satisfaction can
be ensured exactly by construction.
With respect to single-shape accuracy, latent-encodings work well in
specialized scenarios (e.g., large-networks trained on canonically aligned specialized
classes).
With respect to shape-space learning, latent-encodings lie in a simpler
continuous space than weights, which suffer from transposition and reordering
non-injectivities (i.e., multiple weight vectors represent the same implicit).
Nevertheless, weight-encodings allow us to faithfully prepare large diverse datasets of
'real-world' shapes into a vectorizeable representation. We have shown this is simply
\emph{not} possible with existing latent-encodings.
We include the full Thingi10k dataset converted to weight-encoded neural
implicits vectors as a data release\footnote{%
\hyperref%
[github.com/u2ni/ICLR2021]
{https://github.com/u2ni/ICLR2021}
}.
This vectorized data is ripe for meta-learning future work. 
Indeed, concurrent work is already exploring this direction \cite{metasdf}.
We hope our consideration of weight-encoded neural implicits as a first-class
shape representation encourages their use in computer graphics, geometry
processing, machine learning, and beyond.

\bibliography{references}

\begin{thebibliography}{29}
\providecommand{\natexlab}[1]{#1}
\providecommand{\url}[1]{\texttt{#1}}
\expandafter\ifx\csname urlstyle\endcsname\relax
  \providecommand{\doi}[1]{doi: #1}\else
  \providecommand{\doi}{doi: \begingroup \urlstyle{rm}\Url}\fi

\bibitem[Abadi et~al.(2015)Abadi, Agarwal, Barham, Brevdo, Chen, Citro,
  Corrado, Davis, Dean, Devin, Ghemawat, Goodfellow, Harp, Irving, Isard, Jia,
  Jozefowicz, Kaiser, Kudlur, Levenberg, Man\'{e}, Monga, Moore, Murray, Olah,
  Schuster, Shlens, Steiner, Sutskever, Talwar, Tucker, Vanhoucke, Vasudevan,
  Vi\'{e}gas, Vinyals, Warden, Wattenberg, Wicke, Yu, and
  Zheng]{tensorflow2015-whitepaper}
Martin Abadi, Ashish Agarwal, Paul Barham, Eugene Brevdo, Zhifeng Chen, Craig
  Citro, Greg~S. Corrado, Andy Davis, Jeffrey Dean, Matthieu Devin, Sanjay
  Ghemawat, Ian Goodfellow, Andrew Harp, Geoffrey Irving, Michael Isard,
  Yangqing Jia, Rafal Jozefowicz, Lukasz Kaiser, Manjunath Kudlur, Josh
  Levenberg, Dan Man\'{e}, Rajat Monga, Sherry Moore, Derek Murray, Chris Olah,
  Mike Schuster, Jonathon Shlens, Benoit Steiner, Ilya Sutskever, Kunal Talwar,
  Paul Tucker, Vincent Vanhoucke, Vijay Vasudevan, Fernanda Vi\'{e}gas, Oriol
  Vinyals, Pete Warden, Martin Wattenberg, Martin Wicke, Yuan Yu, and Xiaoqiang
  Zheng.
\newblock {TensorFlow}: Large-scale machine learning on heterogeneous systems,
  2015.
\newblock URL \url{http://tensorflow.org/}.
\newblock Software available from tensorflow.org.

\bibitem[Atzmon \& Lipman(2020{\natexlab{a}})Atzmon and Lipman]{SAL}
Matan Atzmon and Yaron Lipman.
\newblock Sal: Sign agnostic learning of shapes from raw data.
\newblock In \emph{IEEE/CVF Conference on Computer Vision and Pattern
  Recognition (CVPR)}, June 2020{\natexlab{a}}.

\bibitem[Atzmon \& Lipman(2020{\natexlab{b}})Atzmon and Lipman]{SALpp}
Matan Atzmon and Yaron Lipman.
\newblock Sal++: Sign agnostic learning with derivatives, 2020{\natexlab{b}}.

\bibitem[Barill et~al.(2018)Barill, Dickson, Schmidt, Levin, and
  Jacobson]{Barill:FW:2018}
Gavin Barill, Neil Dickson, Ryan Schmidt, David~I.W. Levin, and Alec Jacobson.
\newblock Fast winding numbers for soups and clouds.
\newblock \emph{ACM Transactions on Graphics}, 2018.

\bibitem[Chang et~al.(2015)Chang, Funkhouser, Guibas, Hanrahan, Huang, Li,
  Savarese, Savva, Song, Su, et~al.]{chang2015shapenet}
Angel~X Chang, Thomas Funkhouser, Leonidas Guibas, Pat Hanrahan, Qixing Huang,
  Zimo Li, Silvio Savarese, Manolis Savva, Shuran Song, Hao Su, et~al.
\newblock Shapenet: An information-rich 3d model repository.
\newblock \emph{arXiv preprint arXiv:1512.03012}, 2015.

\bibitem[Chen \& Zhang(2019)Chen and Zhang]{chen2018implicit_decoder}
Zhiqin Chen and Hao Zhang.
\newblock Learning implicit fields for generative shape modeling.
\newblock \emph{Proceedings of IEEE Conference on Computer Vision and Pattern
  Recognition (CVPR)}, 2019.

\bibitem[Garland \& Heckbert(1997)Garland and Heckbert]{garland1997surface}
Michael Garland and Paul~S Heckbert.
\newblock Surface simplification using quadric error metrics.
\newblock In \emph{Proceedings of the 24th annual conference on Computer
  graphics and interactive techniques}, pp.\  209--216. ACM
  Press/Addison-Wesley Publishing Co., 1997.

\bibitem[Gropp et~al.(2020)Gropp, Yariv, Haim, Atzmon, and
  Lipman]{gropp2020implicit}
Amos Gropp, Lior Yariv, Niv Haim, Matan Atzmon, and Yaron Lipman.
\newblock Implicit geometric regularization for learning shapes, 2020.

\bibitem[Hart(1996)]{hart1996sphere}
John~C Hart.
\newblock Sphere tracing: A geometric method for the antialiased ray tracing of
  implicit surfaces.
\newblock \emph{The Visual Computer}, 12\penalty0 (10):\penalty0 527--545,
  1996.

\bibitem[Hornik et~al.(1989)Hornik, Stinchcombe, and
  White]{hornik1989mlpuniversl}
Kurt Hornik, Maxwell Stinchcombe, and Halbert White.
\newblock Multilayer feedforward networks are universal approximators.
\newblock \emph{Neural networks}, 2\penalty0 (5):\penalty0 359--366, 1989.

\bibitem[Jacobson et~al.(2013)Jacobson, Kavan, and
  Sorkine-Hornung]{jacobson2013robust}
Alec Jacobson, Ladislav Kavan, and Olga Sorkine-Hornung.
\newblock Robust inside-outside segmentation using generalized winding numbers.
\newblock \emph{ACM Transactions on Graphics (TOG)}, 32\penalty0 (4):\penalty0
  33, 2013.

\bibitem[Jacobson et~al.(2016)Jacobson, Panozzo, Sch{\"u}ller, Diamanti, Zhou,
  Pietroni, et~al.]{jacobson2016libigl}
Alec Jacobson, Daniele Panozzo, C~Sch{\"u}ller, Olga Diamanti, Qingnan Zhou,
  N~Pietroni, et~al.
\newblock libigl: A simple c++ geometry processing library, 2016.

\bibitem[Kahn \& Harris(1951)Kahn and Harris]{kahn1951}
H.~Kahn and T.E. Harris.
\newblock Estimation of particle transmission by random sampling.
\newblock \emph{National Bureau of Standards applied mathematics series}, 12,
  1951.

\bibitem[Kerr et~al.(2018)Kerr, Merrill, Demouth, and Tran]{cutlass}
Andrew Kerr, Duane Merrill, Julien Demouth, and John Tran.
\newblock Cutlass: Fast linear algebra in cuda c, Sep 2018.
\newblock URL \url{https://devblogs.nvidia.com/cutlass-linear-algebra-cuda/}.

\bibitem[Kingma \& Ba(2014)Kingma and Ba]{kingma2014adam}
Diederik~P Kingma and Jimmy Ba.
\newblock Adam: A method for stochastic optimization.
\newblock \emph{arXiv preprint arXiv:1412.6980}, 2014.

\bibitem[Littwin \& Wolf(2019)Littwin and Wolf]{Littwin_2019}
Gidi Littwin and Lior Wolf.
\newblock Deep meta functionals for shape representation.
\newblock \emph{CoRR}, abs/1908.06277, 2019.
\newblock URL \url{http://arxiv.org/abs/1908.06277}.

\bibitem[Liu et~al.(2020)Liu, Zhang, Peng, Shi, Pollefeys, and
  Cui]{liu2019dist}
Shaohui Liu, Yinda Zhang, Songyou Peng, Boxin Shi, Marc Pollefeys, and Zhaopeng
  Cui.
\newblock Dist: Rendering deep implicit signed distance function with
  differentiable sphere tracing.
\newblock In \emph{IEEE Conference on Computer Vision and Pattern Recognition
  (CVPR)}, 2020.

\bibitem[Maturana \& Scherer(2015)Maturana and Scherer]{Maturana2015VoxNet}
Daniel Maturana and Sebastian Scherer.
\newblock Voxnet: A 3d convolutional neural network for real-time object
  recognition.
\newblock In \emph{Ieee/rsj International Conference on Intelligent Robots and
  Systems}, pp.\  922--928, 2015.

\bibitem[Mescheder et~al.(2019)Mescheder, Oechsle, Niemeyer, Nowozin, and
  Geiger]{OccupancyNet}
Lars Mescheder, Michael Oechsle, Michael Niemeyer, Sebastian Nowozin, and
  Andreas Geiger.
\newblock Occupancy networks: Learning 3d reconstruction in function space.
\newblock In \emph{Proceedings IEEE Conf. on Computer Vision and Pattern
  Recognition (CVPR)}, 2019.

\bibitem[Mildenhall et~al.(2020)Mildenhall, Srinivasan, Tancik, Barron,
  Ramamoorthi, and Ng]{NERF}
Ben Mildenhall, Pratul~P. Srinivasan, Matthew Tancik, Jonathan~T. Barron, Ravi
  Ramamoorthi, and Ren Ng.
\newblock Nerf: Representing scenes as neural radiance fields for view
  synthesis.
\newblock In \emph{ECCV}, 2020.

\bibitem[Ohtake et~al.(2005)Ohtake, Belyaev, Alexa, Turk, and
  Seidel]{ohtake2005multi}
Yutaka Ohtake, Alexander Belyaev, Marc Alexa, Greg Turk, and Hans-Peter Seidel.
\newblock Multi-level partition of unity implicits.
\newblock In \emph{Acm Siggraph 2005 Courses}, pp.\  173--es. 2005.

\bibitem[Park et~al.(2019)Park, Florence, Straub, Newcombe, and
  Lovegrove]{Park_deepsdf}
Jeong~Joon Park, Peter Florence, Julian Straub, Richard Newcombe, and Steven
  Lovegrove.
\newblock Deepsdf: Learning continuous signed distance functions for shape
  representation.
\newblock In \emph{The IEEE Conference on Computer Vision and Pattern
  Recognition (CVPR)}, June 2019.

\bibitem[Rahaman et~al.(2019)Rahaman, Baratin, Arpit, Draxler, Lin, Hamprecht,
  Bengio, and Courville]{rahaman2019spectral}
Nasim Rahaman, Aristide Baratin, Devansh Arpit, Felix Draxler, Min Lin, Fred
  Hamprecht, Yoshua Bengio, and Aaron Courville.
\newblock On the spectral bias of neural networks.
\newblock In \emph{International Conference on Machine Learning}, pp.\
  5301--5310. PMLR, 2019.

\bibitem[Sitzmann et~al.(2020{\natexlab{a}})Sitzmann, Chan, Tucker, Snavely,
  and Wetzstein]{metasdf}
Vincent Sitzmann, Eric~R. Chan, Richard Tucker, Noah Snavely, and Gordon
  Wetzstein.
\newblock Metasdf: Meta-learning signed distance functions.
\newblock In \emph{arXiv}, 2020{\natexlab{a}}.

\bibitem[Sitzmann et~al.(2020{\natexlab{b}})Sitzmann, Martel, Bergman, Lindell,
  and Wetzstein]{SIREN}
Vincent Sitzmann, Julien~N.P. Martel, Alexander~W. Bergman, David~B. Lindell,
  and Gordon Wetzstein.
\newblock Implicit neural representations with periodic activation functions.
\newblock In \emph{arXiv}, 2020{\natexlab{b}}.

\bibitem[Tancik et~al.(2020)Tancik, Srinivasan, Mildenhall, Fridovich-Keil,
  Raghavan, Singhal, Ramamoorthi, Barron, and Ng]{tancik2020fourier}
Matthew Tancik, Pratul~P Srinivasan, Ben Mildenhall, Sara Fridovich-Keil,
  Nithin Raghavan, Utkarsh Singhal, Ravi Ramamoorthi, Jonathan~T Barron, and
  Ren Ng.
\newblock Fourier features let networks learn high frequency functions in low
  dimensional domains.
\newblock \emph{arXiv preprint arXiv:2006.10739}, 2020.

\bibitem[Wang et~al.(2018)Wang, Sun, Liu, and Tong]{Wang-2018-AOCNN}
Peng-Shuai Wang, Chun-Yu Sun, Yang Liu, and Xin Tong.
\newblock {Adaptive O-CNN: A Patch-based Deep Representation of 3D Shapes}.
\newblock \emph{ACM Transactions on Graphics (SIGGRAPH Asia)}, 37\penalty0 (6),
  2018.

\bibitem[Xu et~al.(2020)Xu, Fan, Yuan, and Singh]{Xu2020}
Yifan Xu, Tianqi Fan, Yi~Yuan, and Gurprit Singh.
\newblock Ladybird: Quasi-monte carlo sampling for deep implicit field based 3d
  reconstruction with symmetry.
\newblock In \emph{Proc. {ECCV}}, 2020.

\bibitem[Zhou \& Jacobson(2016)Zhou and Jacobson]{zhou2016thingi10k}
Qingnan Zhou and Alec Jacobson.
\newblock Thingi10k: A dataset of 10,000 3d-printing models.
\newblock \emph{arXiv preprint arXiv:1605.04797}, 2016.

\end{thebibliography}
\bibliographystyle{iclr2021_conference}

\appendix
\section{Appendix}
\subsection{The \repName\ File Format}

Our compact weight-encoded neural implicit is designed to be effortlessly
consumed and integrated into existing graphics and geometry processing
pipelines.
For each trained model, the chosen architecture and similarity transformaton
matrix (since all geometries are normalized to the unit sphere) are written as
the first bytes before encoding the learned weights $\theta$ into an HDF5 format file. 

For a fixed architecture, the instructions to evaluate the estimated SDF is the
same for any point \emph{and} any shape.
This SIMD property allows multiple geometries to be evaluated in
parallel.
The fixed storage profiles and memory layout of our learned implicit functions
provide consistent query and rendering speeds.
We store our model weights using the HDF5 format. This allows easy integration into Tensorflow (below) which can load our model natively. We additionally support the loading of arbitrary weight-encoded neural implicit through the "High Five" HDF5 C++ library (https://github.com/BlueBrain/HighFive) for rendering and meshing. 

\begin{lstlisting}[language=Python]
	import tensorflow as tf
	import numpy as np
	
	# load model "key" dictating architecture. SIMD.
	sdfModel = tf.keras.models.model_from_json(open('key.json'))
	
	# load specific weight for Standford bunny geometry
	sdfModel.load_weights('bunny.h5')
	
	# generate 128x128x128 grid for SDF queries
	K = np.linspace(-1.0,1.0,128)
	grid = [[x,y,z] for x in K for y in K for z in K]
	
	# infer SDF at each point
	S = sdfModel.predict(grid)
\end{lstlisting}

\subsection{ Error Driven Conversion }

\begin{figure}[htb!]
    \centering
    \includegraphics[width=\linewidth]{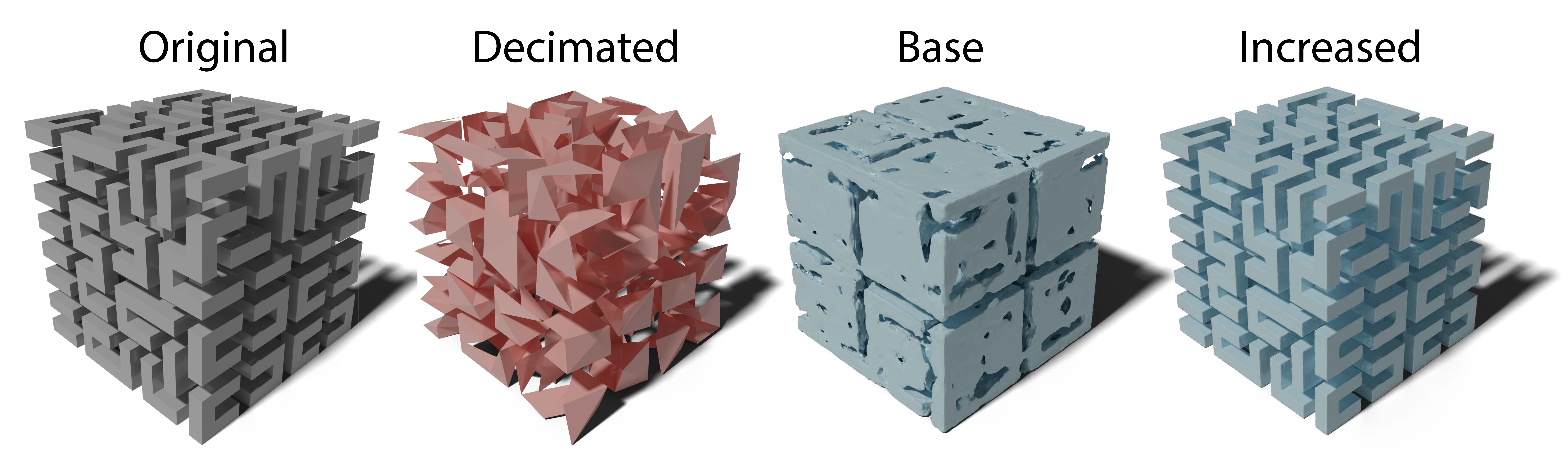}
    \caption{With only 7553 parameters, our \textit{base} \repName\ format can lack the representative power to converge on highly complex geometries (similar to decimated mesh with same memory footprint). Increasing the network capacity to equal the memory impact of the original mesh results in near perfect reconstructions. tbuser (left) under CC BY.}
    \label{fig:failure}
\end{figure}

We fix the architecture during the Thingi10k dataset conversion, resulting in a constant and compact memory footprint. If, however, maintaining a target surface reconstruction quality is of more importance to a fixed memory cost, we can instead shift to an error driven surface fitting approach (much like classical approaches \citep{ohtake2005multi}), scaling network architecture complexity based on the input geometry. As each generated \repName\ encodes its own architecture, such an approach results in smaller architectures for simpler geometries and larger ones for topologically-complex geometries. We visualize the effect of error driven optimization in Figure \ref{fig:failure}, where we perform a simple grid search until reaching a user-desired surface error threshold.

Based on our conversion of the Thingi10k dataset, we find that a majority of models are well represented using our base configuration (Fig. \ref{fig:reconstructionHistograms}) -- if desired, geometries that fall within the tails of the complexity distribution can be retrained with larger architectures, again until we reach a desired surface fidelity. This decision can be further informed by whether SIMD and fixed memory access patterns are beneficial to the underlying application.

\subsection{SIREN and Fourier Features} \label{highFreq}

In an effort to improve the reconstruction quality of our weight-encoded neural implicits we explored recent work focused on improving MLPs ability to represent high frequency signals. We experimented with three methods: namely we investigated using the SIREN \citep{SIREN} activations, positional encoding \citep{NERF}, and Fourier features \citep{tancik2020fourier}. Each of these approaches have lead to impressive resuts for high-fidelity reconstructions of 3D surfaces mitagating the known problem that MLPs learn low frequency signals faster \citep{rahaman2019spectral}. 

\citet{NERF} define their positional encodings as, 
\begin{equation}
	\gamma(p) = (sin(2^0 \pi p), cos(2^0 \pi p) ,..., sin(2^{L-1} \pi p), cos(2^{L-1} \pi p))
\end{equation}
where $\gamma$ is a mapping from $\R$ into the higher dimensional space $\R^{2L}$. 

While \citet{tancik2020fourier} expands on this approach with random gaussian features yielding the mapping function, 

\begin{equation}
	\lambda(p) = (cos(2\pi\ \textbf{B} p), sin(2 \pi \textbf{B} p)) 
\end{equation}

where each entry in $\textbf{B} ∈ \R^{m×d}$ is sampled from  $\mathcal{N} (0, \sigma^2
)$, and $\sigma$ is left as a hyperparameter specific to each problem.

\begin{figure}[htb!]
	\includegraphics[width=1.0\linewidth]{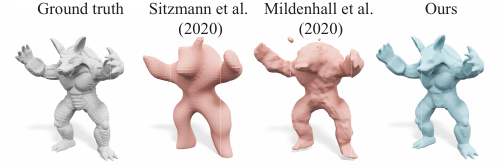}
	\vspace{-20pt}
	\caption{Results of using \cite{NERF} and \cite{SIREN}} 
	\vspace{-10pt}
	\label{fig:sirenandff}
\end{figure}

We evaluate both of these approaches by mapping each axis (x,y,z) of our sampled points to the higher dimensional space. We find that when the network architecture is of sufficient width these mappings work exceptionally well. We evaluated using $\gamma$ with various $L$ configurations ranging from 4 to 10. Unfortunatly, we find that our light weight (and intentionally underparamerterized) architecture struggles to learn from the augmented input signal. We visualize the affect of positional encodings when $L=10$ in Figure \ref{fig:sirenandff}. Similarly, we see drastic degredation of quality when employing $\lambda$ for mapping to a default embedding size of 256 (not shown as we were unable to march). These approaches are clearly practical methods for reconstructing high-fidelity surfaces, but with our focus on minimizing the number of parameters the cost of mapping to a higher dimension input is too high. 

Our experimental setup for evaluating \citet{SIREN} periodic activation consisted of modifying an existing tensorflow \citep{tensorflow2015-whitepaper} implementation to accept our spatial queries as input and signed distances as target. We train the SIREN model to 200 epochs with a learning rate of $5e^{-5}$ and the same loss as our own configuration. Interestingly, we find that the SIREN model produces smoother approximations of the armadillo's surface (see Figure \ref{fig:sirenandff}) but lacks fine detail. Once again, when increasing our model complexity to just 8 layers of 64 hidden units, we start to see the benefits of the periodic activation yielding much better approximations of the surface then our relu activation. For our base configuration of just 7553 parameters we choose to continue using RELU activation, but where high-fidelity weight-encoding neural implicits are required, SIREN should be employed.

\subsection{ Representation Compactness }
\begin{figure}[htb!]
    \includegraphics[width=0.9\linewidth]{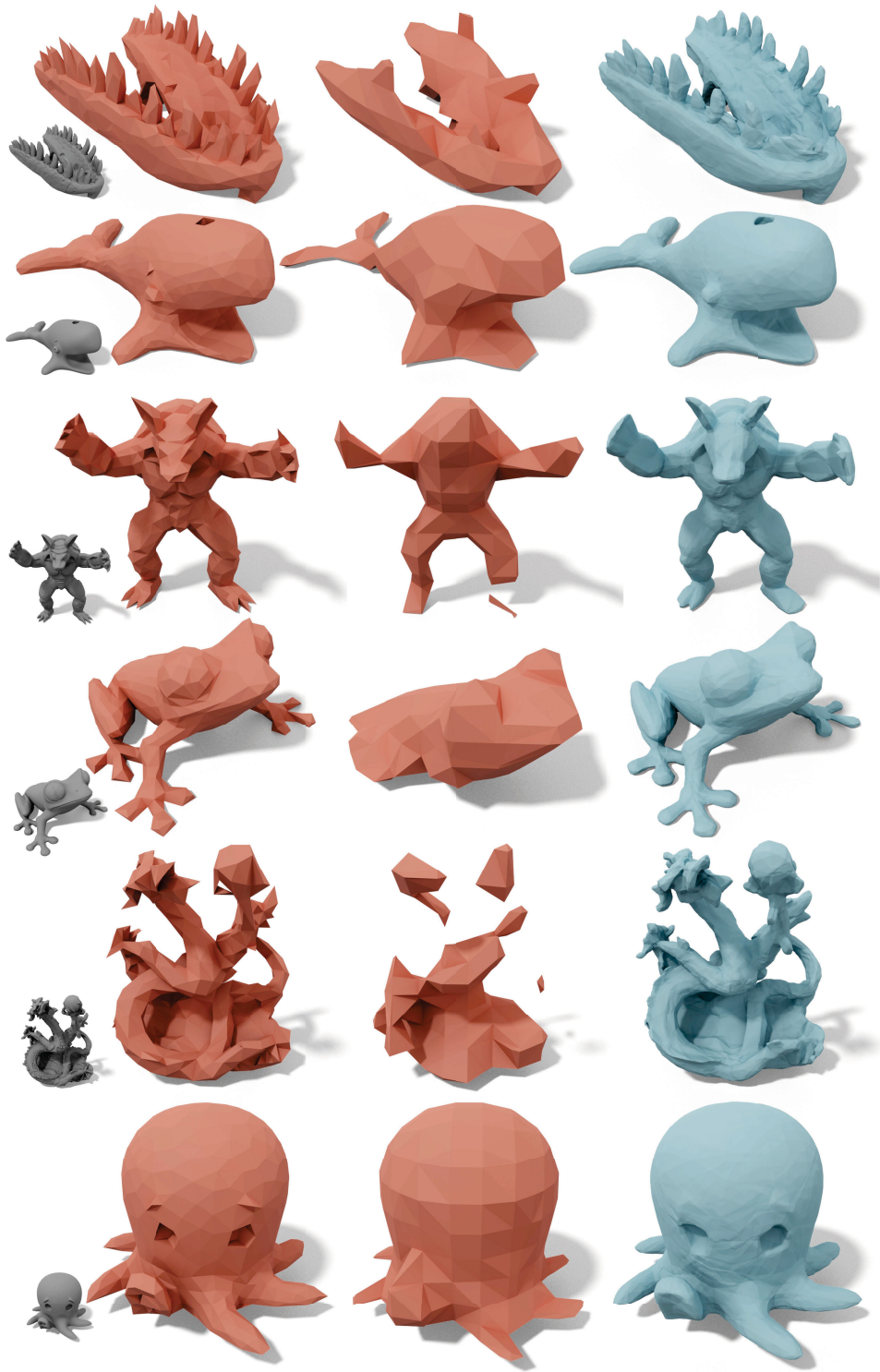}
    %\vspace{-20pt}
    \caption{\label{fig:decimation}Our learnt \repName\ format (right) can be shown to better approximate the original surface (grey, inset) compared to adaptive decimation of the original triangle mesh \cite{garland1997surface} (left) and uniform signed distance grid (middle) with equal memory impact. gpvillamil (skull), Makerbot (whale), morenaP (frog), artec3d (dragon), JuliaTruchsess(octopus) under CC BY.} 
    %\vspace{-10pt}
    \label{fig:decimated}
\end{figure}

\end{document}